# ADQUEX: Adaptive Processing of Federated Queries over Linked Data based on Tuple Routing

Amin Beiranvand
Department of Electrical and Computer Engineering,
Isfahan University of Technology
a.beiranvand@ec.iut.ac.ir

Nasser Ghadiri[1]
Department of Electrical and Computer Engineering,
Isfahan University of Technology
nghadiri@cc.iut.ac.ir

**Abstract:** *Recent achievements of linked data implementations and the increased number of datasets available on the web as linked data, has given rise to the need and tendency toward processing federated queries over these datasets. Due to the distribution of linked data across the web, the methods that process federated queries through a distributed approach are more attractive to the users and have gained more prosperity. In distributed processing of federated queries, we need methods and procedures to execute the query in an optimal manner. Most of the existing methods perform the optimization task based on some statistical information, whereas the query processor does not have precise statistical information about their properties, since the data sources are autonomous. When precise statistics are not available, the possibility of wrong estimations will highly increase, and may lead to inefficient execution of the query at runtime. Another problem of the existing methods is that in the optimization phase, they assume that runtime conditions of query execution are stable, while the environment in which federated queries are executed over linked data is dynamic and non-predictable. By considering these two problems, there is a great potential for exploiting the federated query processing techniques in an adaptive manner. In this paper, an adaptive method is proposed for processing federated queries over linked data, based on the concept of routing the tuples. The proposed method, named ADQUEX, is able to execute the query effectively without any prior statistical information. This method can change the query execution plan at runtime so that less intermediate results are produced. It can also adapt the execution plan to new situation if unpredicted network latencies arise. Extensive evaluation of our method by running real queries over well-known linked datasets shows very good results especially for complex queries.*

**Keywords:** *Processing Federated Queries over Linked Data, Adaptive Query Processing, Federation of SPARQL Endpoints*

[1] *Corresponding author. Address: Department of Electrical and Computer Engineering, Isfahan University of Technology, Isfahan, Iran. Phone : +98-311-391-9058, Fax: +98-311-391-2450, Alternate email: nghadiri@gmail.com*

# 1 Introduction

With the extensive publishing of data on the web as linked data [1], the web of data is growing concurrently with the syntactic web. Linking data that is distributed across the web requires a standard mechanism to define the existence and semantics of relationships between the entities described by this data [2]. This mechanism is the RDF framework [3]. Entities in RDF are defined as triples. A triple consists of a subject, a predicate and an object. An RDF graph is a set of subject-predicate-object triples and a set of RDF graphs is called a dataset[2]. The SPARQL [4] query language is a W3C recommendation, [4] and is used to query RDF graphs. The SPARQL queries are based on *triple patterns*, or simply *patterns*. A set of patterns is called a Basic Graph Pattern (BGP) [4]. The difference between triple patterns and RDF triples is that in triple patterns, we can put variables instead of the subject, predicate or object of a triple. A BGP matches with a subgraph of an RDF graph if its nodes or vertices can replace the variables of the given BGP. This subgraph, if found, is the result of the query execution [4].

The Linked Open Data (LOD) is a collaborative project aimed at publishing open datasets as linked data. Currently the LOD cloud contains about 1014 datasets [5], that belong to different domains and are published by different individuals. Given the fact that %56.11 of these datasets have links to at least one other dataset [5], a potentially great demand exists for running federated queries over linked data. Federated SPARQL queries are those queries that integrate data from multiple datasets distributed across the web. In the federated queries, each triple pattern can be matched with the subgraphs of different datasets. If a subgraph of a dataset matches a triple pattern, the dataset is called a *source* for the triple pattern. A triple pattern may have zero or multiple sources.

Two general approaches exist for running federated queries over linked data. The first approach is to copy linked data sets distributed across the web to a data warehouse, followed by running the query against this central repository. In the second approach, called distributed query processing, data remains at its original location and a mediator runs the query. The mediator receives the query, sends each part of the query to the corresponding data source for execution, and integrates the returned results (the tuples resulted from execution of subqueries), and eventually returns the final result (the tuples passed through all query operators and processed by them). The performance of the data warehouse approach is generally higher than the distributed approach, but serious challenges arise, such as keeping the warehouse up to date, making its deployment a difficult task. The data warehouse architecture is also incompatible with inherent distribution of the linked data. Therefore, most recent methods for executing the federated queries follow the distributed approach.

[2] In this paper, we use the terms *dataset* and *data source* interchangeably. The term *tuple* is also used in database literature. We used it to describe our systems as some routing concepts stem from the database domain.

The mediator must find a good plan for optimal execution of the query. An important parameter that the mediators try to decrease is the number of the *intermediate results* (the tuples produced by each operator during the query execution and passed to the next operator). It helps in lowering the number of tuples that need to be processed, hence minimizing the cost of query execution. The costs of different plans are estimated based on a cost estimation model. This model needs some information about the properties of the datasets. One of the main challenges in the federated query processing for linked data is that we have no precise information about the features of their data, due to the independence and autonomy of the datasets. Some methods rely on some presorted statistical information by the datasets to find the optimal query plan [6-8], whilst only %13.46 of datasets provide meta-data about themselves [5]. Other methods are also proposed [9, 10] that use heuristic techniques to find the optimal plan, but they are subject to errors. Another challenge in this context is that the query execution environment is dynamic. The parameters such as the response time of the data sources, the network latencies, and the amount of available memory are subject to change, which in turn affects the execution time of the queries.

Given the two aforementioned challenges, it seems that exploiting the *adaptive* query processing techniques [11, 12] that can change the query plan according to real runtime variables, will help us to improve federated query processing over linked data. In our proposed method, we have extended the Eddy operator [13]. Eddy is an adaptive operator to be used in the linked data context and we use it for query execution. The proposed method does not require any statistical information for finding the optimal plan. Actually it finds the optimal plan at runtime. It will change the query execution plan at runtime if it does not perform well enough, without any change in the correctness of the query results. The most important features of the proposed method are as follows:

- On-demand processing of the federated SPARQL queries
- No need to statistical information about the data sources
- Adapting the plan at runtime so that less intermediate results are produced
- Adapting the plan at runtime to cope with the network latencies, so as to preventing the query processor from blocking and the query execution goes on without disruption

The proposed method is extensively evaluated by applying it to some well-known datasets and benchmarks. The results are interpreted to investigate its behavior in different situations.

The rest of the paper is organized as follows. An overview of the related work is given in section 2. Section 3 describes our proposed ADQUEX (ADaptive QUery EXecutor) architecture for adaptive execution of the federated SPARQL queries, after a brief overview of the basic ideas. The experimental results with several scenarios in different configurations are presented in Section 4. Section 5 concludes the paper and points out to future research.

## 2 Related work

Processing of the SPARQL queries over RDF has been an active research area in recent years. A few survey papers have been published that neatly categorize the existing approaches in this domain [14, 15].

Current methods for processing of SPARQL queries over RDF can be classified under two categories. The first group of methods aim at offering a solution for storing and querying RDF data at a central location, such as Virtuoso [16], RDF-3X [17] and SW-Store[18].

The second group follow the federated query processing approach over linked data distributed across the web. Our focus in this paper is on the second group of systems.

Federated query processing techniques over linked data are also categorized by Tran and Ladwig [19] into three groups, in terms of how they discover the sources that are associated with the queries. The first group assumes that the sources are strictly specified in the query. Our proposed system fits into this group. Systems in the second group do not require the sources to be strictly specified in the query [20-22]. They use information about the sources at compile time to determine the sources associated with the query. The required information are either supplied by the sources themselves (for example by using void [23]), or the query processor acquires and stores information in a data catalog for subsequent use. Federation engines that are based on distributed query processing in databases fall into the first group or the second group. The third group of systems discover the sources associated with the queries at runtime based on the content that they observe. The link traversal methods [24-26] are among the methods of this group. In the link traversal methods, a preliminary set of sources are discovered through existing URIs inside queries, and some portions of data are acquired and inspected during query execution to discover other related sources. This will be repeated until no more sources are found [27]. A hybrid method is also proposed by Tran and Ladwig for query execution that uses information from compile time and tries to discover new sources at runtime [19]. It also performs source ranking at runtime. As we mentioned earlier in this section, our proposed method assumes that all related sources of the query are identified at compile time. In the following, we overview the systems that are in the same category as our system. Among the methods described below, ARQ and SPARQL-DQP [28] fit into the first category just like our system, as they require the sources to be declared in the query, and the rest of the methods fit into the second category, as they perform source-selection automatically.

ARQ is a SPARQL query execution engine for Jena[3]. ARQ supports the SERVICE keyword, and it can execute federated SPARQL queries that are defined using the SERVICE keyword.

[3] http://jena.apache.org

DARQ is a system that requires the data sources to provide some information about their predicates, the number of their triples, and number of triples of a predicate, the selectivity of a triple pattern as well as other parameters using service descriptions. Selection of the sources in DARQ is performed using such statistical information. The statistical information are also used in the query optimization phase for finding the best join order. A problem with DARQ is that it only supports the queries in which triple predicates are bound. Another problem is that only a few number of data sources actually provide service descriptions, and we cannot rely on this incomplete information for source selection and query plan optimization [7].

FedX is another system that does not require the data sources to provide any data catalog or metadata. For selecting the source of a pattern, the pattern is sent to every data source through an ASK query and if any source responded by YES, the source will be selected for that pattern. Three methods are used in this system for optimizing a query: (1) join ordering, (2) exclusive groups, and (3) bind joins. This system performs join ordering based on a heuristic variable counting cost estimation method. Two operators namely Nested Loop Join (NLJ) and bind join are used in FedX for the physical join. FedX is not an adaptive federation engine and is a static system. If any network latencies happen at runtime for the data sources or if the order of joins is not optimized according to runtime information, it will not change the plan at runtime [10, 29].

In the SPLENDID system it is assumed that the data sources provide their statistical information such as the number of triples and the number of unique predicates, subjects and objects through void descriptions. It uses this information for source selection and for optimizing the query plan. Three operations are performed for query optimization by SPENDID federation engine to execute the queries more efficiently: query rewriting, forming exclusive groups and finding the optimized order of joins. For implementing the joins, two physical operators of hash join and bind join are used. SPLENDID also does not change the query plan at runtime and according to different situations. Therefore, it is not an adaptive federation engine [8].

The ADERIS federation engine does not expect prior statistical information from data sources. It gets the required information through the SPARQL queries. The statistics gathering phase is performed before running the queries. ADERIS builds a predicate table for each triple pattern and after running the triples by the data sources, the results are materialized in the predicate tables. Then it joins the predicate tables to find the query results. The predicate table is a table with two fields for subject and object and the value of a predicate determines the name of the table. A useful feature of ADERIS is that it can change the join algorithm at runtime. If the execution of a triple pattern takes a long time to complete, it will change the predicate table to a non-materialized table. Therefore, it can be classified as an adaptive federation engine [30, 31].

The SemWIQ federation engine gathers statistical information that it receives through SPARQL queries in a data catalog. Then it uses this information for source selection and query optimization. Due to the specific methods that SemWIQ uses for source selection, it can execute some restricted types of queries. One of this restrictions is that the subject of all triples must be a variable and the type of every subject has to be defined by `rdf:type` predicate [6].

SPARQL-DQP supports SPARQL 1.1 queries. It has proposed and implemented a collection of query rewriting rules for decreasing the query processing time [28].

ANAPSID first acquires a list of predicates of a data source through sending a SPARQL query to the SPARQL endpoints of that query and performs source selection based on this list [32]. Then it performs query decomposition and optimization based on heuristic methods presented by Montoya et.al [9]. This federation engine has extended the Xjoin physical join operation [33] and exploits it for performing joins. ANAPSID can adapt itself to low memory situations by moving the tuples from the main memory into the disk storage. When no effective work can be done due to the network latencies, it will process the tuples that were moved to the disk to perform effective work and generate results.

Among the aforementioned federation engines, ANAPSID [32] and ADERIS [30] are adaptive. Other methods are not adaptive and try to find the query plan at compile time. However, not much information is available about the properties of data sources at compile time since they are autonomous. Estimations that are made at compile time for finding the optimal query plan may diverge from reality. On the contrary, our ADQUEX method does not require statistical information about the data sources and is able to find a good plan for query execution based on the information it acquires at runtime. ADQUEX is also different from ANAPSID [32] and ADERIS [30] since it tries to change the order of the operators in the plan, according to the runtime conditions.

# 3 The proposed ADQUEX system

In this section, our proposed system is introduced. The system is based on the Eddy query processor [13]. In normal operation of a pipelined plan, every operator processes a single tuple and sends it directly to the next operator which is specified in the plan. The idea behind Eddy is that after each operator processes a tuple, it sends the tuple to a router, instead of sending to another operator. The router then sends the tuple to another operator. The main point here is that the next operator is not necessarily the next operator as defined in the plan; it can be selected from the operators that are able to process that tuple. This allows the router to have multiple choices when selecting the destination operator and the router should select the operator that makes it possible to run the query in a more efficient way.

Indeed, the Eddy system is a router [13]. It processes the queries by routing the tuples toward query operators. As shown in Figure 1, the tuples are entered from the relations into Eddy, which sends them to the operators. Each operator processes a tuple and sends the generated tuples back to Eddy if any partial results are generated by processing that tuple. To prevent infinite loops and sending the tuples to output eventually, Eddy assigns a *Done* status to each tuple. This status determines which operators have already processed the tuple. Through handling this status, Eddy is able to conceive the number of operations that have processed the tuple, and it will send a tuple to output when it is processed by all of the operations. The *Done* status cannot be just a number. For any given tuple, Eddy must determine exactly which operators it has been passed through. The reason for this control is that Eddy should not send any tuple more than once to each operator and this rule is implemented by the *Done* status.

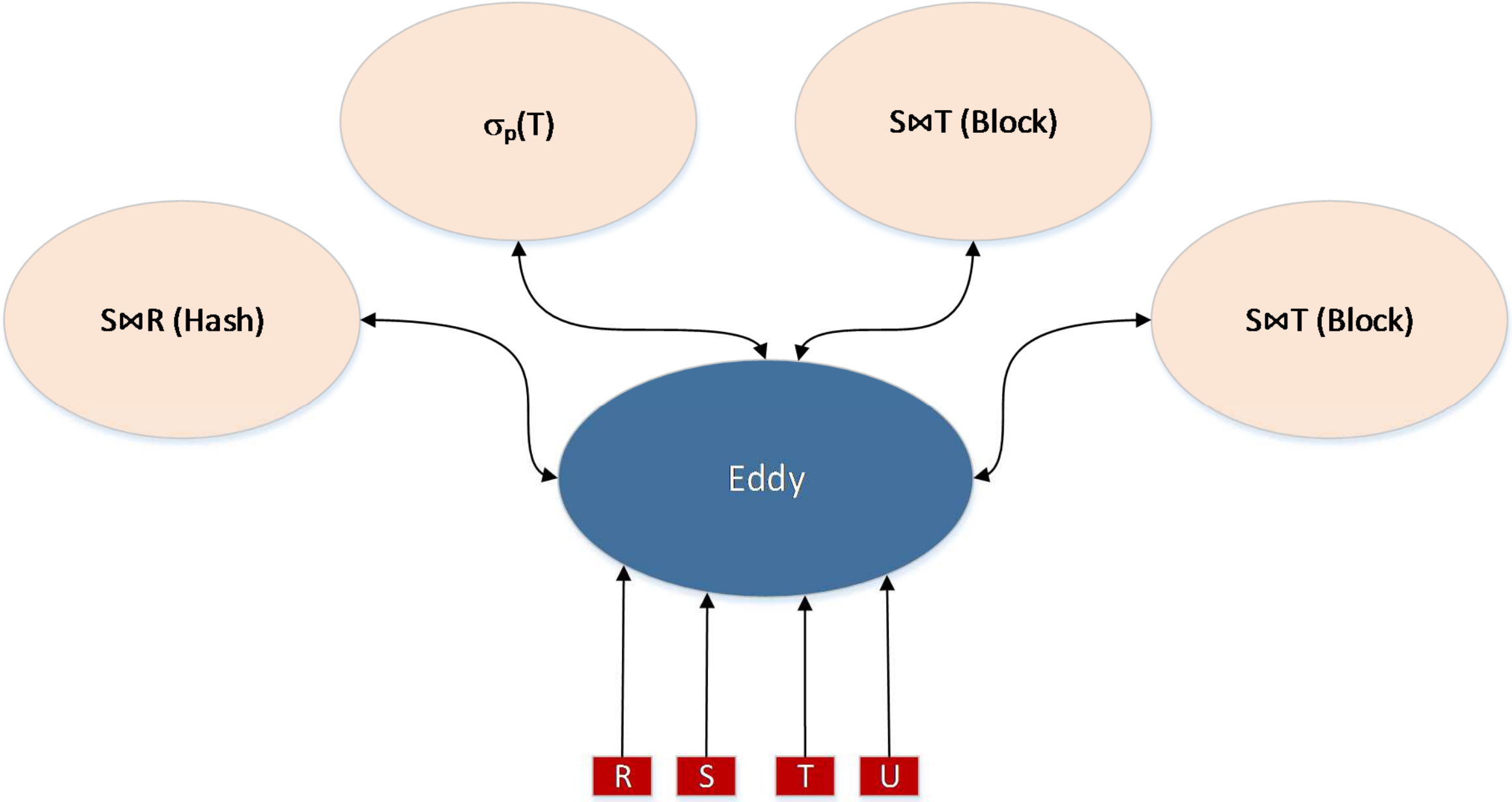

**Figure 1:** Using Eddy in a pipeline [13]

Eddy can modify the execution plan by changing the order of sending the tuples to the operators. Eddy is very flexible and it can modify the execution plan on a per tuple basis. Making decision about the order of sending a tuple to different operators is guided by a routing strategy. The routing strategy is able to determine the next operation in order to minimize the query execution cost. This decision is based on parameters like the cost of an operation, eligibility of an operation, network delays and so on. A pipelined plan is depicted in Figure 2. A tuple from R cannot be sent to the join operator number 2 (since the tuples in R do not have property b), unless they are joined with a tuple from S, and then they are sent to the second operator. While the tuples in S can be joined with both the first join operator and the second join operator.

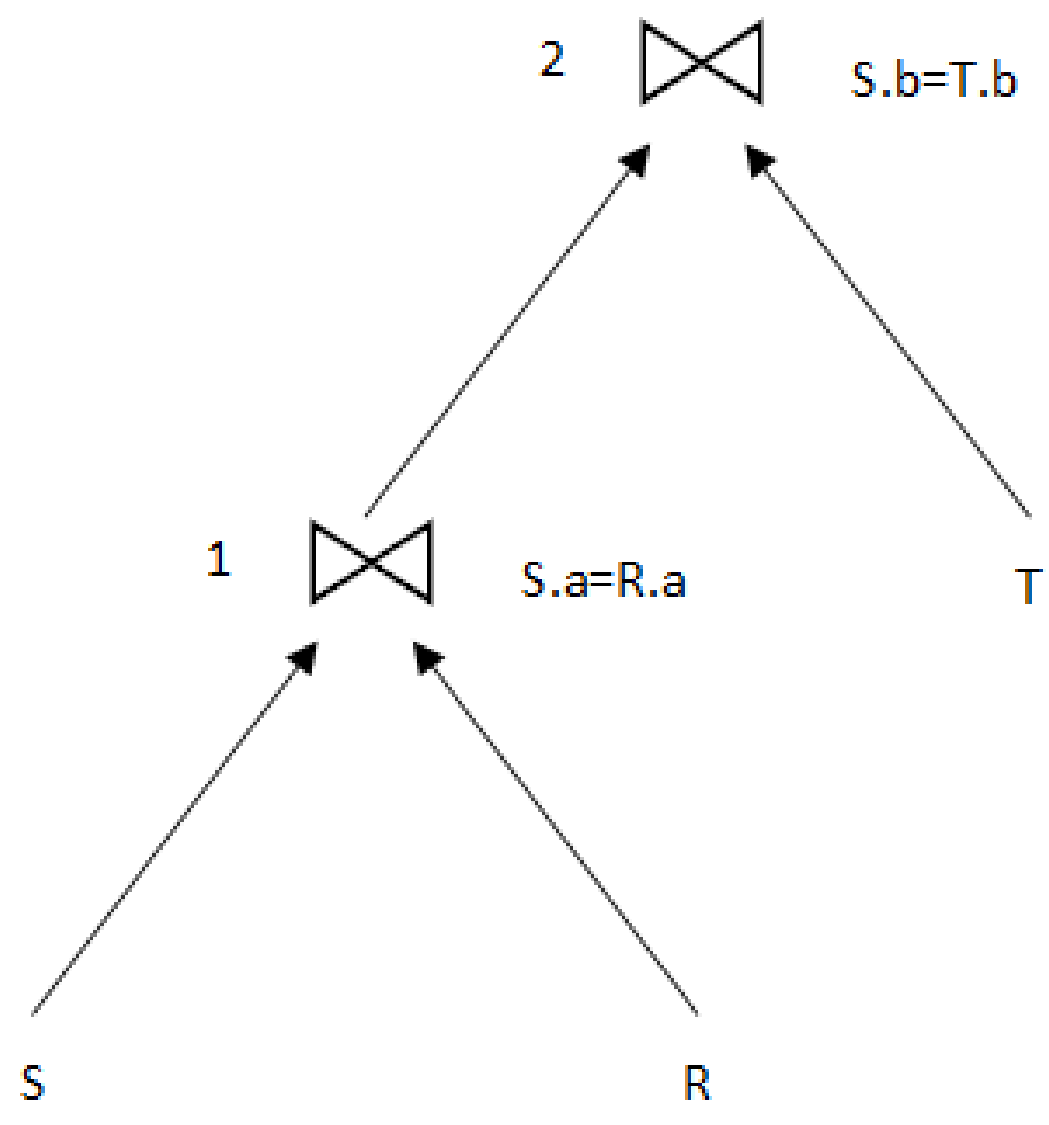

**Figure 2:** A pipeline plan, each operator sends the tuple after processing to the next operator

To determine which operators a tuple can be sent to and to prevent sending tuples to the wrong operator, Eddy assigns a *Ready* status to each tuple, which determines to which operators this tuple can be sent, and to which should not be sent.

The environment for processing of the federated SPARQL queries on SPARQL endpoints is unpredictable, and we have no precise information about the parameters like the operator selectivity and the cardinality of sub-queries. Moreover, the underlying network parameters such as latencies are also unpredictable. Therefore, federated SPARQL queries will potentially have a better performance if they are executed *adaptively*. We have exploited the idea of executing queries through routing of tuples [13] for adaptive execution of queries in our system which is described in the following subsection. The notations that we will use throughout the paper are summarized in Table 1.

**Table 1:** Notations and symbols

| *Symbol* | *Description* |
|---|---|
| $C$ | The cost of a join operator, calculated by Algorithm 1 |
| $D$ | The amount of network latency that has no effect on the query plan and can be ignored |
| $I$ | Number of input tuples to a join operator |
| $F$ | Coefficient for the effect of network latency on the cost of a join operator |
| $O$ | Number of output tuples from a join operator |
| $Op_x$ | Operator number $x$ |
| $P(i)$ | Priority of tuple $i$ |
| $Q_x$ | Query number $x$ |
| $S_x$ | Sub-query number $x$ |
| $T(i)$ | The number assigned to the $i^{th}$ tuple by a sub-query executor |

### 3.1 The architecture of ADQUEX

The architecture of our system is shown in Figure 3. The input to the system is a SPARQL 1.1 query where the address of SPARQL endpoints are specified using the SERVICE keyword. Our system supports SPARQL version 1.1 federated queries. The system currently does not support SPARQL 1.0 queries and it requires to exactly define which part of the query is associated with which SPARQL endpoint.

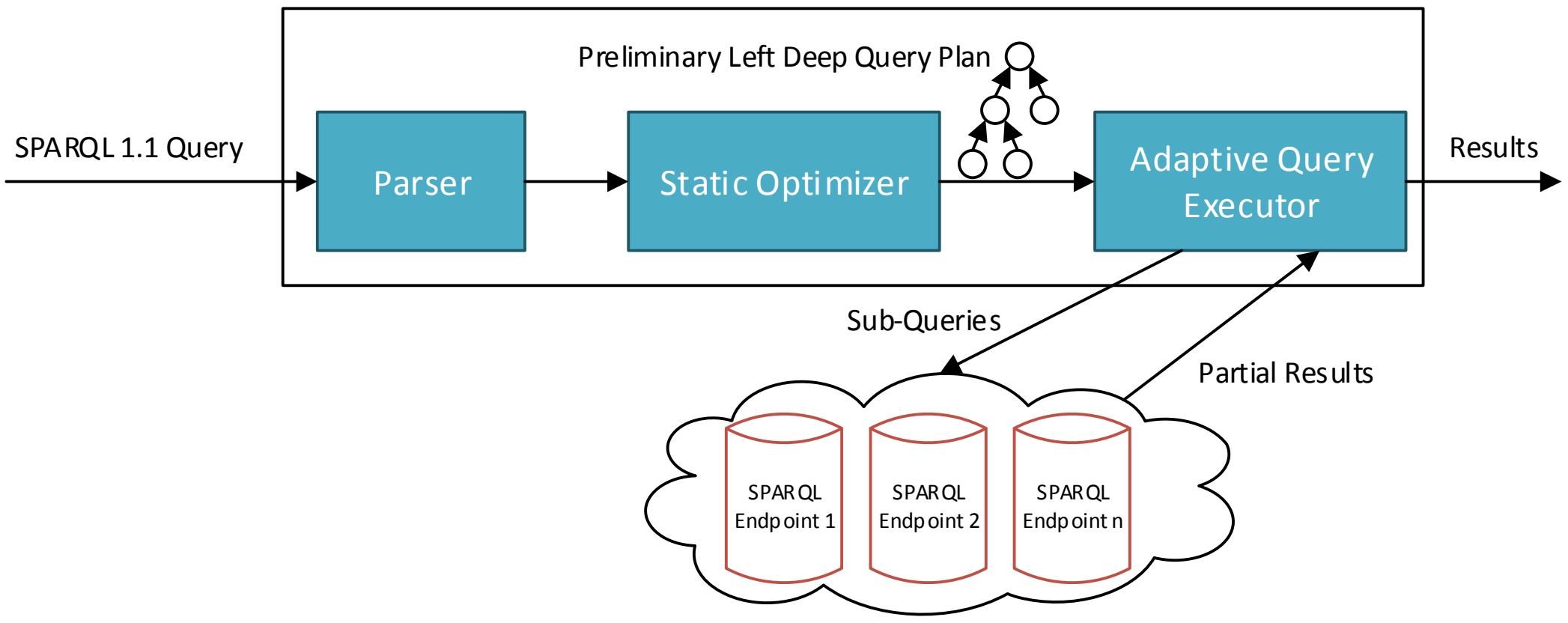

**Figure 3:** The proposed ADQUEX architecture

The role of each component in our system is as follows:

1) *Parser*: This component checks and validates the syntax of query and converts it to an execution plan if there are no syntax errors. An execution plan is a tree that specifies the order of executing the query operators.
2) *Static optimizer*: At this step, we optimize the query using *filter pushing* and *projection pushing* [34] as two query rewriting methods. It should be noted that we have no prior statistical information such as the cardinality of sub-queries or the selectivity of joins for optimizing the query. Indeed, our system the optimization of join orders is performed by the query executor and by using runtime information only. This is a unique feature of our system that it does not require any initial information about the datasets for optimized execution of a federated query. The reason for not using statistical information is that in real-world situations, we have usually not much information about the datasets such as cardinality and selectivity of join operators. Moreover, even if we had such statistics, they were subject to change at runtime so that the statistics would become invalid, and our plan would not be optimized anymore. The static optimizer eventually builds a preliminary plan and passes it to the query executor for execution. The plan built by this query optimizer is called *initial plan*, as it will be used merely for initiating the query execution. This plan may be modified several times during the execution for the purpose of optimization.

3) *Query executor*: This component is responsible for executing the query in an adaptive manner. Our focus in this paper is on this component of the system. As shown in Figure 4, the query executor itself consists of several components as described in the next section.

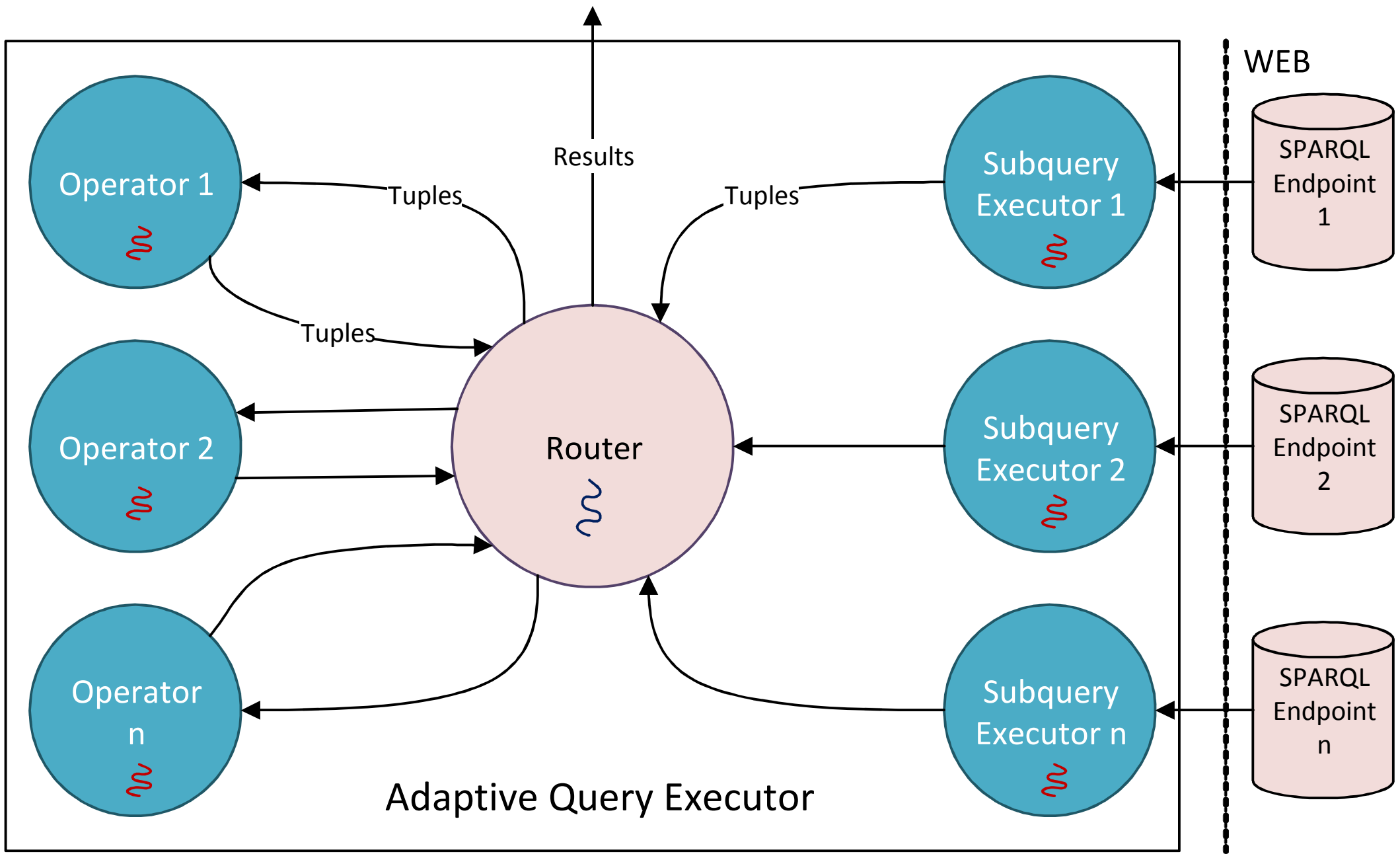

**Figure 4:** Components of the adaptive query executor

## *3.2 The router*

This component is responsible for routing the incoming tuples. The router has a priority queue in which the tuples are buffered. Sub-query executors put the tuples in this queue, and the router fetches the tuples and send them to appropriate operators. Hereafter, we call this queue the *router priority queue*. The router priority queue is implemented as a blocking queue which causes the router to enter a blocked state when there are no more tuples in the queue. It will wait until a new tuple enters the queue.

Every operator also has a queue which functions as a buffer. This is a FIFO queue and is implemented as a blocking queue. The router has access to the operator queues and sends the tuples to the operators by putting them in operator queues. This will prevent missing the tuples when an operator is processing a tuple and the next tuple enters the queue. Moreover, when an operator has no tuples to process, it will be blocked to save CPU time.

If an operator wants to send a tuple to the router, it can put the tuple in the router priority queue almost in same way that sub-query executors do.

## 3.3 *The routing strategy*

The router sends the tuples to the operators based on a routing strategy. In this section, we describe the routing strategy that we used to route the tuples.

In our routing strategy, we focused on two parameters: the number of intermediate results and the network latencies. We aim to route the tuples such that the intermediate results are decreased and, at the same time, perform the routing in a way that network latencies do not cause the query execution to halt. For example, consider the query plan shown in Figure 5. Suppose that we send *S1* tuples to the *?film* join first, and then to the *?director* join since it produces fewer intermediate results. At runtime, we observe that *S2* is facing a delay and it does not send any tuples. In this case, our system will change the plan, and will send the *S1* tuples to the *?director* join first and then to the *?film* join. In this case, the newer plan is better since if we continue the preliminary plan, the tuples of *S1* which go to the *?film* join will not join with *S2* tuples as *S2* tuples arrive in with a delay. The join is actually performed when the tuples of *S2* arrive and the result of the *?film* join must be sent to the *?director* join afterwards. During this period, no effective processing can be performed. On the contrary, executing the second plan makes it possible to perform some useful processing until the tuples arrive at *S2*, i.e. the join between tuples from *S1* and *S3*.

```
select ?dbdirector ?directorpage ?title
where{
        SERVICE <http://192.168.1.2:8890/sparql>{          S1
            ?film a dbpedia-owl:Film.
            ?film dbpedia-owl:director ?director.}
        SERVICE <http://192.168.1.3:8890/sparql>{          S2
            ?imdbfilm owl:sameAs ?film.
            ?imdbfilm dc:title ?title}
        SERVICE <http://192.168.1.4:8890/sparql>{          S3
                ?concept owl:sameAs ?director.
                ?concept nytimes:topicPage ?directorpage.}
}
```

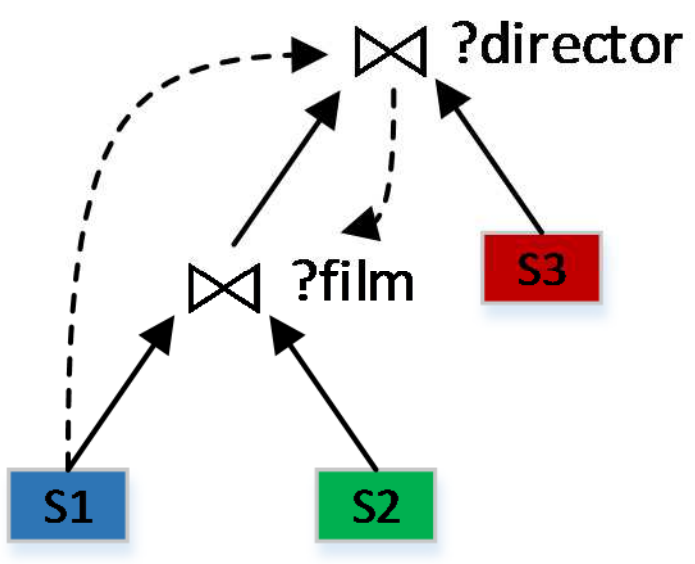

**Figure 5:** A SPARQL query and its plan. The preliminary plan is shown with arrows.

We use Algorithm 1 to estimate the cost of a join operator. When we want to route a tuple, we will send it to the operator which has the lower cost based on this algorithm.

**Algorithm 1:** Pseudo-code for estimating the cost of a join operator

```
EstimateCost (joinId)
{
        BSD ← RuntimeStatistics.getBaseSourceDelay(joinId)
        O ← RuntimeStatistics.getCountOutput(joinId)
        I ← RuntimeStatistics.getCountInput(joinId)
        If (BSD > Parameters.D)
                C ← (O/I)+Parameters.F
        Else
                C ← (O/I)
        Return C
}
```

In Algorithm 1, *I* is the number of tuples that enter an operator and *O* is the number of output tuples of an operator which are returned to the router. The *RuntimeStatistics* is a singleton object that is signaled by the router when it sends a tuple to a join operator to increase the number of input tuples for that join operator. Also, when a join operator produces a result, the *RuntimeStatistics* is signaled to increase the number of outputs for that join operator. In this way, the input and output statistics of all joins are kept by *RuntimeStatistics* and will be used when routing the tuples to select the best destination join for any given tuple. The ratio of *O/I* actually determines the selectivity of an operator. If a tuple can be sent to more than one operator, ignoring the network latency, that tuple will be sent to the operator with lower *O/I* ratio. It will decrease the number of intermediate results. However, we are not done completely and by taking the network latencies into account, we have to change our decision making method as we describe.

In this algorithm, *D* is the maximum latency that can be tolerated with no cost assigned to it and the system does not respond to network latency. The preliminary plan that is given to the router is a deep left tree. Every join operator in this plan has two inputs. At least one input is connected to a basic sub-query. A basic sub-query is a sub-query that must be sent to the SPARQL endpoint of its corresponding data source for execution. The task of each sub-query executor is to run a basic sub-query. Now if a sub-query executor has not sent a tuple to the router during a time threshold of *D*, then an extra cost of *F* is considered for the join operator that is connected to that basic sub-query. The *F* parameter actually defines the intensity of response to network latency. If *F* is low, then our goal will tend to decreasing the number of intermediate results. Alternatively, if *F* is high, it means that we are more interested in performing effective work during network latencies, even if the intermediate results are increased. For example, suppose that we have two operators *Op1* and *Op2* with C(*Op1*)=0.2 and C(*Op2*)=0.6, neglecting the latencies. The basic sub-query connected to *Op1* has latency and *F*=0.7. Now if we are able to send a tuple to both *Op1* and *Op2*, the router will send it to *Op2* since by taking network latency into account, the cost of *Op1* will be about 0.9 which is higher than the cost of *Op2*. However, we have generated more intermediate results by this selection as the ratio of output/input is higher for *Op2*. If the parameter *F* was small then *Op1* would be preferred to *Op2* even if we had network latencies, thus generating less intermediate results.

## *3.4 Sub-query executors*

Every sub-query executor is responsible for sending a sub-query to its associated SPARQL endpoint to be executed by that SPARQL endpoint. When tuples are returned by the SPARQL endpoint as a result of sub-query execution, the sub-query executor should receive and prepare them and send them to the router afterwards.

Before sending a tuple to the router, two status labels *Done* and *Ready* must be attached to it. As mentioned earlier in this section, the *Ready* status shows that to which operators a tuple can be sent and the *Done* status shows that to which operators a tuple has been already sent to. We implemented these status labels as two sets of bits. The number of bits in each status depends on the number of operators in the query plan. For instance, if we have 5 operators, then we need 5 bits for the *Ready* status and 5 bits for the *Done* status. Every bit represents a single operation and can be set to 0 or 1 accordingly. All bits of the *Done* status for all tuples that are generated by the query executors have a value of zero, as they have not been processed by any operator yet. On the contrary, the bits of the *Ready* status must be determined according to the preliminary query plan. All tuples that are resulted from the execution of the same basic sub-query have exactly the same values of *Ready* bits, since they have the same variables and thereupon they can be sent to the same operators. Therefore, in order to decrease overheads and to prevent unnecessary computation of these bits for every tuple, we implemented sub-query executors in such a way that they only get the value of the *Ready* bits through the preliminary plan once. Afterwards, they preserve the values and assign them to all of the tuples. In addition to *Ready* and *Done* bit sets, the sub-query executor must assign a priority to every tuple, which attached as a number to each tuple. In Figure 6 the structure of the query tuples of Figure 5 are shown. One can observe that each tuple has a *Binding Set* section where the values of tuple variables are stored.

| | Priority | Done | Ready | Binding Set |
|---|---|---|---|---|
| S1 Tuples | 0 | 00 | 11 | ?film ?director |
| S2 Tuples | 0 | 00 | 10 | ?film ?imdbfilm ?title |
| S3 Tuples | 0 | 00 | 01 | ?concept ?director ?directorpage |

**Figure 6:** The structure of query tuples of Figure 5, each tuple has two states (*Done* and *Ready*), a priority and a set of values

In some federated SPARQL queries, a sub-query may have more than one related source. Since our systems does not perform source selection, the SPARQL endpoints of such a sub-query must be specified in the query. In a normal situation where every sub-query has a single related source, this can be achieved through the SERVICE keyword. However, when there are multiple related sources, it is impossible to specify the addresses of several SPARQL endpoints in one SERVICE keyword. A solution in this case is to write a sub-query for each source, using the SERVICE keyword, and to integrate the results of these sub-queries using the UNION operator. When our system receives such a sub-query, in a way similar to normal queries, a sub-query executor is built for each sub-query. The tuples are fetched from the related source and sent to the router. The router sends them to the proper operators according to the *Ready* bits. The point here is that the *Ready* bits of the tuples that belong to the same sub-query, but come from different sources, are equal.

The order of putting tuples in the router's priority queue can affect the number of intermediate results. In the query shown in Figure 5, if we send the tuples of *S1* to ?*director* join first, less intermediate results will be generated. Now suppose that all tuples of S1 are ordered and put in the beginning part of the priority queue. The router fetches a few tuples from the queue and sends them to the operators, but when the costs of the operators are computed, the costs of both operators will be zero. This is due to the fact that all tuples which are sent to the operators have been *S1* tuples, and since no tuples other than *S1* tuples are sent to any operator, no join is performed and the output of the operators becomes zero, hence the zero cost of the operators. This is an unrealistic cost, as the tuples of *S1* may be sent to *?film* operator which leads to increasing the number of intermediate results. The real cost of the operators is determined when the tuples from sub-queries other than *S1* are also sent to the operators. Supposing that we have already routed all tuples of *S1*, this computation of cost will no longer help, as the flexibility only exists in routing the tuples of *S1*, and there is no flexibility for routing other tuples. To prevent this from happening, and to distribute the tuples uniformly in the router's priority queue, we use Eq.(1) for assigning the priority values, *P(i)*, to the tuples, instead of assigning zero priority to every tuple.

$$P(i) = Random(maxRandom) - T(i) \qquad (1)$$

In Eq.(1) *T(i)* is the number of a tuple in a sub-query. By assigning the priorities through Eq.(1), the tuples will be put in the router's priority queue without being ordered, preventing a large number of tuples from a single sub-query to be put in the queue one after another. Moreover, the tuples arrived earlier will be put in the queue with higher priority. By changing the value of *maxRandom* we can control the degree to which this rule is applied. All basic tuples do have a priority value of less than or equal to *maxRandom*, but the tuples resulted from joins must have a priority value higher than *maxRandom* to be put above of others in the queue and to be processed earlier. Later in Section 4, we will evaluate and represent the effect of assigning priorities to the tuples using this method.

## *3.5 The physical join operator*

We perform the join operation using the physical operator Symmetric Hash Join (SHJ) [35, 36]. There are several reasons for using the SHJ algorithm. Firstly, SHJ has frequent moments of symmetry [13]. Moments of symmetry are the time periods that we can change the query plan, without invalidating the state. Secondly, SHJ is not a blocking operator, which makes it very suitable for a pipelined plan and the pipelined plan whose join operators are SHJ is a fully pipelined plan. The SHJ operator is quite applicable to the environments like the world-wide web where the tuples arrive in an interleaved manner, since it does not depend on the order of inputs. It can process every tuple when it receives the tuple from any of its inputs. The classic hash join has only one hash table. Only one of the inputs are built and the other input is probed to it. The weakness of the classic hash join is that it requires all the build-side inputs to arrive before the probing can be started. This method is not suitable for the environments like the web, since the operator must be blocked to completely receive an input. Unlike the hash join operator, the SHJ operator builds a hash for every input. When a tuple arrives, it will be stored in a proper hash table and probed to the opposite table. This will enable the SHJ to process data from any of its inputs depending on availability.

Figure 7 depicts the status of *?film* operator of the query which was already shown in Figure 5. A hash table stores the tuples of *S1* and another hash table stores the tuples of *S2*. In Figure 7(a) the status of two hash tables of *?film* join operator at the time of *t* is shown. Observe that the *S1* hash table can store both *S1* basic tuples and the tuples resulted from the join of *S1* tuples and tuples from other sub-queries (*S3* in this example). At the time of *t+1* when the tuple (*film3*, $S2_u$) is sent by the router to this join operator, the operator first determines that which hash table should be used for inserting this tuple. Here the tuple includes some variables from the *S2* sub-query, so it should be inserted into the hash table of the *S2* tuples.

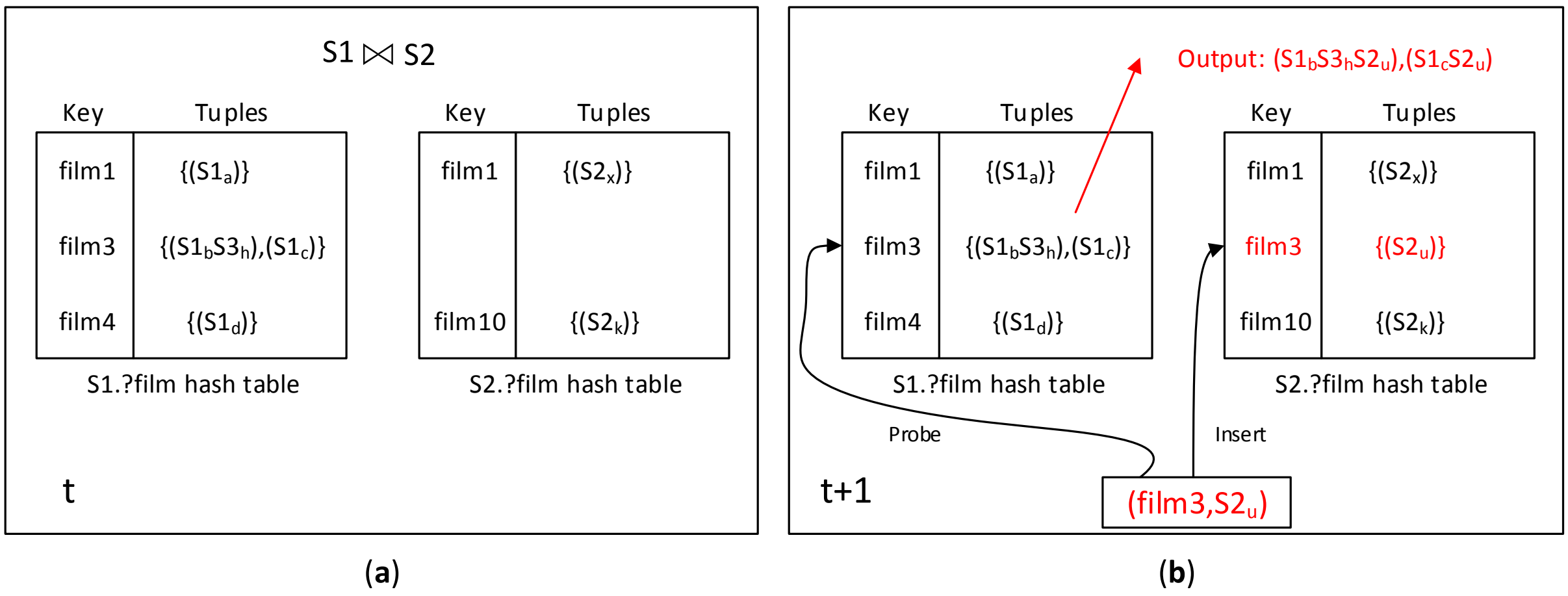

**Figure 7:** The state of operator for joining *S1,S2*: (a) at the time of *t*, (b) at the time of *t+1*.

After inserting this new tuple in the hash table of the *S2* tuples, the operator will probe it in its opposite table where the new tuple having a *film3* key is matched with two tuples having *film3* key in the hash table of the *S1* tuples. Now we can join the matched tuples with the probed tuple. Eventually, two tuples are generated after joining, as can be seen in Figure 7(b).

The join operator should send the tuples resulted from the join to the router again. However, the status bit sets of *Ready* and *Done* as well as the priority of the tuples resulted from join must be exactly set before sending to the router, to prevent invalid routing of tuples by the router. After this stage, the operator can put the tuples in the router's priority queue to be forwarded to other operators again, or to be sent to the final output.

## *3.6 Multiple sub-queries with the same join operator*

In SPARQL queries, it is possible to join multiple SERVICEs with the same join operator. For example, consider a simple query as shown in Figure 8. It contains three sub-queries that are joined by the *?person* join operator.

```
Select * where
{
SERVICE <http://192.168.1.2:8890/sparql> { ?person a dbpedia-owl:person. }
SERVICE < http://192.168.1.3:8890/sparql > {?lmdbperson owl:sameAs ?person.}
SERVICE < http://192.168.1.4:8890/sparql > {?nyperson owl:sameAs ?person.}
}
```

**Figure 8. A sample query having three sub-queries with the same join operator**

The initial plan for this query is shown in Figure 9. As described earlier, since the operators of both join operations in this query are equal, and the router can send any tuple from each of the sub-queries to Join 1 as well as Join 2. In other words, the two *Ready* bits are true for all tuples of these two sub-queries (there are two Ready bits if we consider the join operators only).

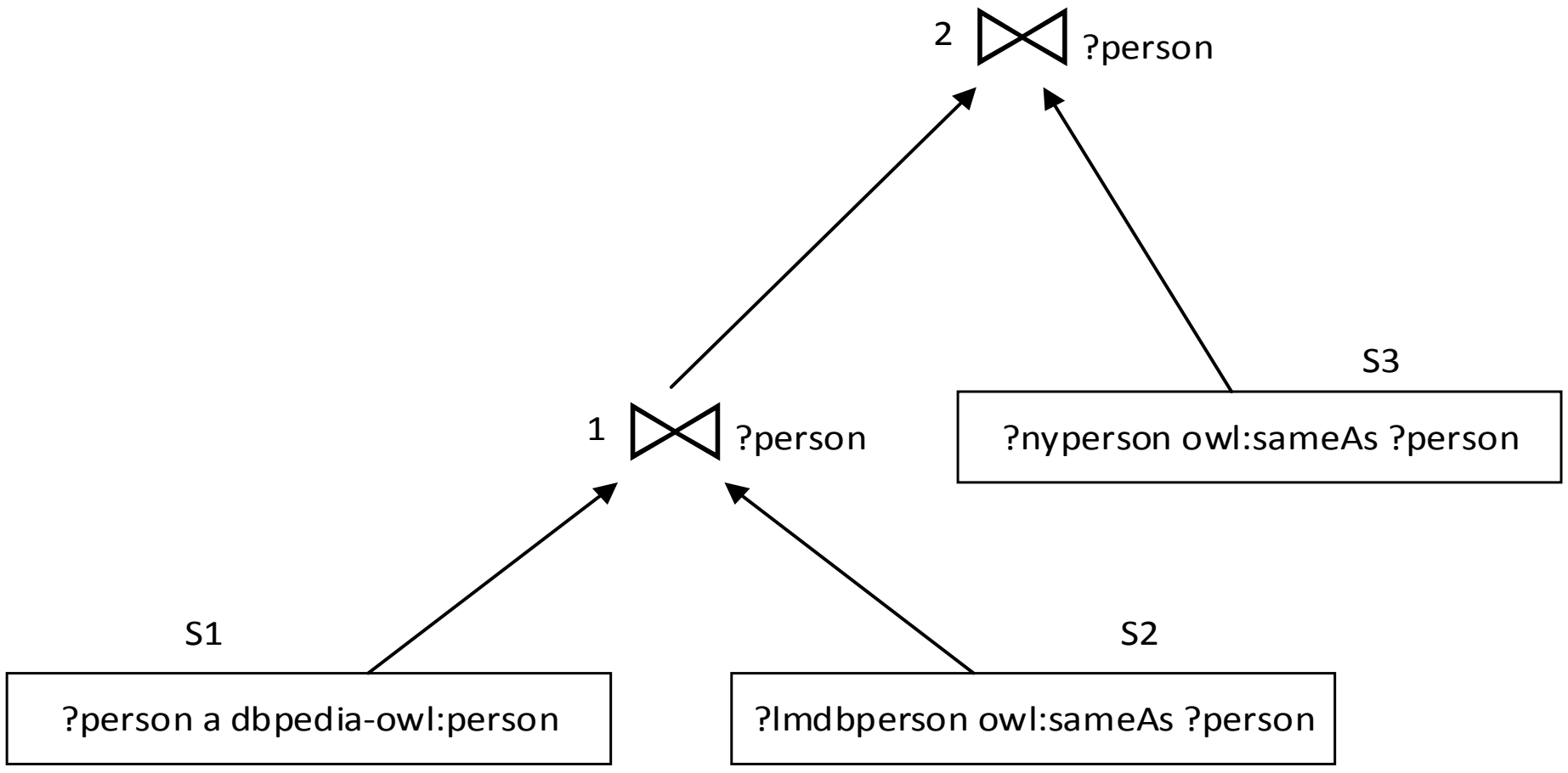

**Figure 9. Initial plan for the sample query shown in Figure 8**

However, in this case where all tuples can be sent by the router to any of the join operators regardless of the sub-query they belong to, and the physical operator SHJ is used, the results of the query may be incomplete. Some tuples of the result may not be returned in this case. For example, suppose that we have three tuples {?person=Brad Pitt}, {?lmdbperson=lmdb Brad Pitt, ?person=Brad Pitt} and {?nyperson=ny Brad Pitt, ?person=Brad Pitt} in the queue and they are sent by the router to the operators at times *t*, *t+1* and *t+2*. The SHJ operator is implemented in a way that when a tuple is sent to the join operator, if the tuple contains another tuple that was attached to the right side of the join in initial plan, it will be inserted into the right hash table, and probed into the left table. If it contains no such tuple, the hash operation is performed in reverse direction. Now suppose that the router has sent the first two tuples at times *t* and *t+1* to the Join 2. As the sub-queries of these tuples are not attached to the right side of Join 2 in initial plan, they are inserted into the left table of this join, and the probe will not produce any result since the right table is empty. Now since the router is not restricted, if it opts to send the third tuple at time *t+2* to Join 1, the resulting status will be as shown in Figure 10. The execution is ended without returning any results, while we expected a single tuple to be returned.

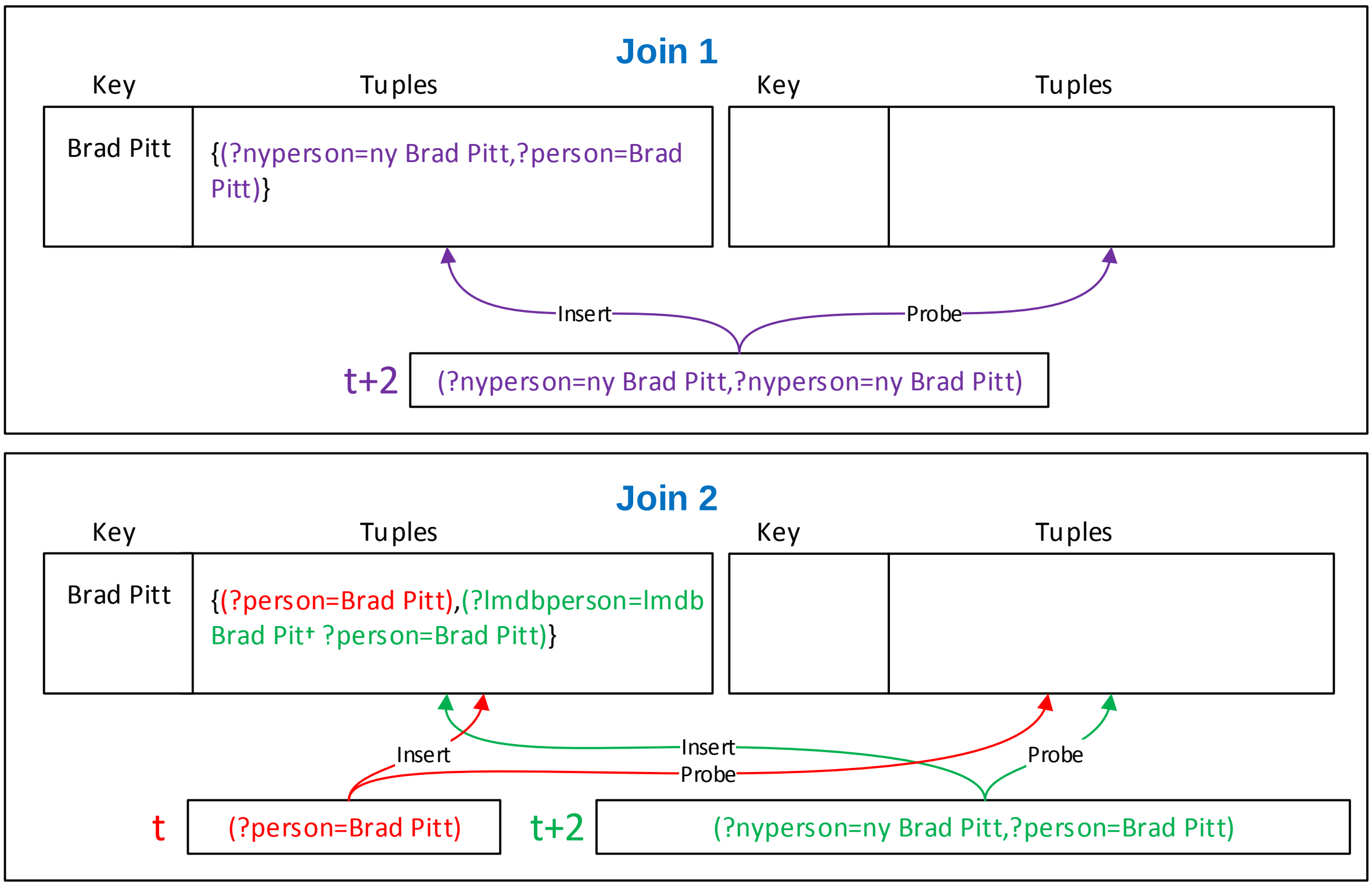

**Figure 10. The status of joins after sending three tuples**

Our proposed method for solving this problem is as follows. When determining the *Ready* bits for the tuples of the sub-queries, we allow only the tuples of one sub-query to select the join deliberately, but we restrict the tuples of other two sub-queries to a single join. For example, in the query shown in Figure 8, we can allow the tuples of the sub-query *S1* to be routed to both Join 1 and Join 2, but we restrict the tuples of *S2* to be routed only to Join 1, and the tuples of *S3* only to Join 2. By using this method, the problem of missing tuples will no longer arise. At the same time, the number of intermediate results  can be decreased by the selection of correct join operator for the tuples that can select any of the two joins independently.

## 4 Experimental results

In this section, the proposed federated SPARQL query routing method is evaluated using real-world datasets, and its performance is compared with existing methods. We implemented our ADQUEX federation engine in Java language version 6. We also used Sesame library in sub-query executors to make connection with the SPARQL endpoints. To evaluate the system we used the datasets in FedBench [37]. The properties of the datasets are shown in Table 2. We loaded each of the datasets into a separate PC node. Each PC has an Intel Core i5 processor running at 3.0GHz, with 4GB of RAM and Ubuntu 14.04 operating system. For loading and running the SPARQL endpoints on each PC we installed the Virtuoso triple store and loaded data into it. The federation engines were installed on a PC with Intel Core i5 processor running at 3.0GHz, with 4GB of RAM and Windows 7 operating system, all connected through a 100Mbps LAN to SPARQL endpoints.

**Table 2:** The properties of datasets, #Triples shows the number of RDF triples in each dataset

| Dataset | Version | #Triples |
|---|---|---|
| DBpedia subset | 3.5.1 | 43.6M |
| NY Times | 2010-01-13 | 335k |
| LinkedMDB | 2010-01-19 | 6.15M |
| Jamendo | 2010-11-25 | 1.05M |
| GeoNames | 2010-10-06 | 108M |
| KEGG | 2010-11-25 | 1.09M |
| Drugbank | 2010-11-25 | 767K |
| ChEBI | 2010-11-25 | 7.33M |

Our evaluation will be presented in three parts. In the first part, the proposed routing strategy is evaluated without considering the network latencies. In the second part, network latencies are also taken into account for evaluating the routing strategy. Finally, our proposed federation engine is compared with similar engines in terms of query execution times.

## *4.1 Evaluation of Routing Strategy without Network Latencies*

In this subsection, we evaluate our proposed routing strategy in ADQUEX against five queries and show that our method can adaptively find an optimal plan. In this set of experiments, the queries are executed on an ideal 100Mbps network with no latency. We compared the number of intermediate results generated by our proposed routing strategy with the number of intermediate results generated by the best plan, worst plan, the Lottery routing strategy [13], and a strategy without assigning priorities to the tuples using Eq.(1), labeled as “Our Strategy-Random”. We measured the number of intermediate results of best plan and worst plan by running all possible plans except the plans that contain a Cartesian product. We executed each query ten times by each method and computed the average of the number of generated intermediate results. For a better comparison, we have also computed the variance of this ten executions in term of the number of generated intermediate results.

The properties of the queries are shown in Table 3. All queries have two datasets and the address of the SPARQL endpoint is specified by SERVICE keyword. The triples that belong to the same SPARQL endpoint were sent as a single group to the SPARQL endpoint, so in the federation engine a single join operation will be performed. However, if we perform a single join, there will be a single plan which all the methods will execute, and the evaluation will not be correct. To increase the number of joins, and for a thorough benchmarking of the routing strategy, we intentionally divided each SERVICE into several SERVICEs for these set of queries.

**Table 3:** The properties of queries used for evaluation of the routing strategy. #P is the number of triples patterns, #J is the number of joins that the federation engine must perform, and #R is the number of query results.

| Query | #P | #J | #R |
|---|---|---|---|
| Q1 | 5 | 3 | 148 |
| Q2 | 7 | 3 | 3 |
| Q3 | 4 | 3 | 1301 |
| Q4 | 5 | 3 | 1620 |
| Q5 | 4 | 2 | 11 |

In Figure 11, different strategies are compared in terms of the average number of intermediate results and Figure 12 shows the variance of the number of intermediate results. As you can see, our method has successfully changed the query plan in an adaptive manner at runtime for all of the queries, so that the number of intermediate results it produced during query execution is very close to the best plan. For the *Q2* query, the difference between the number of intermediate results of our method with those of the best plan is relatively high and it seems that our method has found the best plan somewhat lately. The small difference between the number of intermediate results produced by our method and by the best plan is normal, since it takes some time for the optimum plan to emerge. One can observe that the number of our intermediate results is much lower than the worst plan.

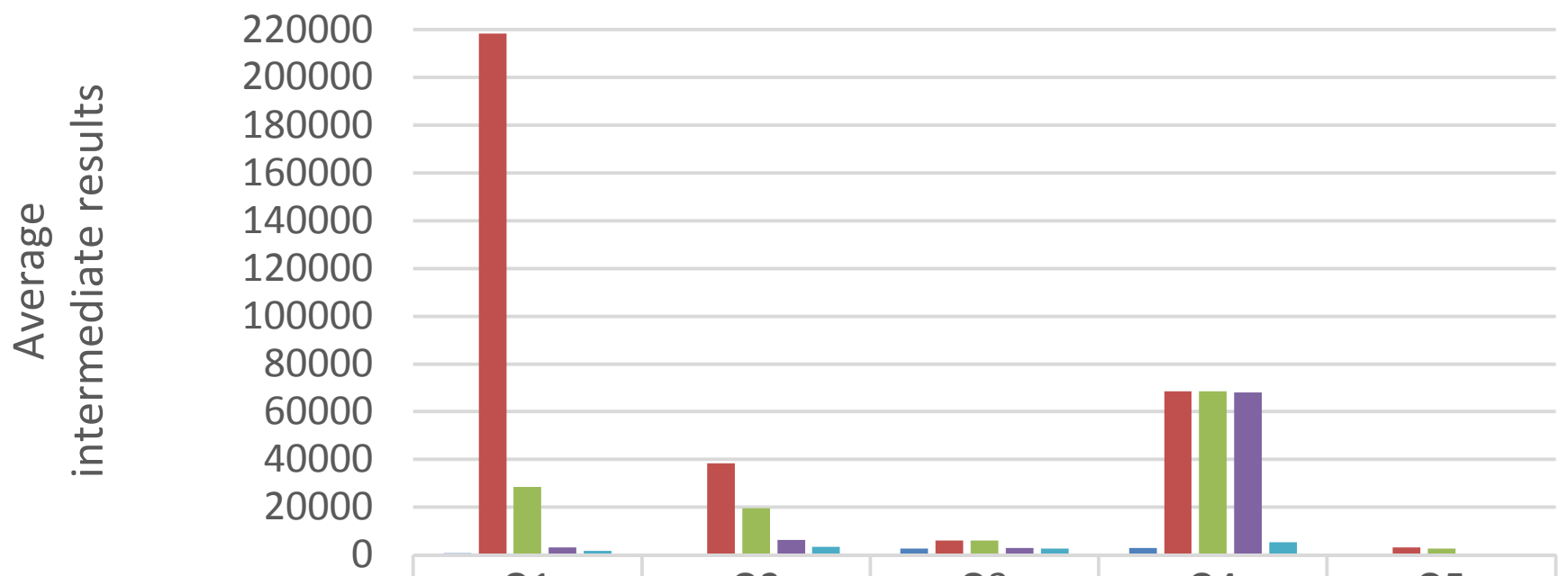

| | Q1 | Q2 | Q3 | Q4 | Q5 |
|---|---|---|---|---|---|
| Best Plan | 695 | 2 | 2602 | 2910 | 11 |
| Worst Plan | 218375 | 38345 | 6075 | 68479 | 3244 |
| Lottery | 28370.9 | 19431.2 | 6073 | 68478.3 | 2554.3 |
| Our Strategy | 3081.9 | 6173.7 | 2943.8 | 68015.4 | 58.3 |
| Our Strategy+Random | 1705.3 | 3462.15 | 2726.1 | 5233 | 173.1 |

**Figure 11:** The average number of generated intermediate results after ten times of query execution with each routing strategy

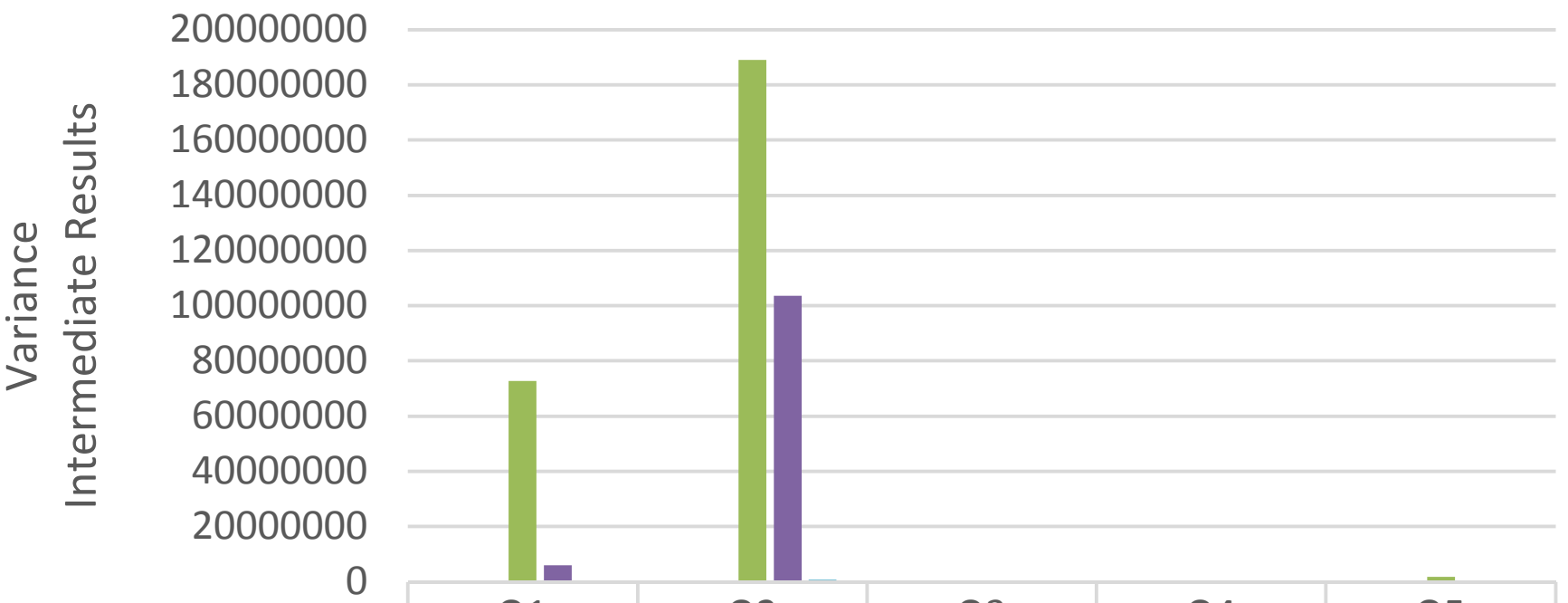

| | Q1 | Q2 | Q3 | Q4 | Q5 |
|---|---|---|---|---|---|
| Best Plan | 0 | 0 | 0 | 0 | 0 |
| Worst Plan | 0 | 0 | 0 | 0 | 0 |
| Lottery | 72833731.9 | 188952026 | 0 | 2.2 | 1807323.3 |
| Our Strategy | 5990400.8 | 103551589 | 32984.8 | 1242.3 | 22372.9 |
| Our Strategy+Random | 567.6 | 737626 | 4062.5 | 9820 | 24.5 |

**Figure 12:** The variance of the number of generated intermediate results after ten times of query execution

Moreover, our proposed method performs much better than the Lottery strategy and the experiments show a significant difference between them. As you can see, our strategy has performed better by producing a lower number of intermediate results, even if we don't assign priorities to the tuples using Eq.(1). However, the variance of the number of intermediate results is not satisfactory and is less stable. We will show that our complete method, which exploits Eq.(1) to assign priorities to the tuples, performs better in terms of the average number of intermediate results (compared to our no-priority method), as well as a much lower variance that makes it a stable method.

## *4.2 Evaluation of Routing Strategy when Facing Network Latencies*

In this subsection, we evaluate our routing strategy by considering network latencies. For this evaluation, we used the *Q1* query as shown in Figure 13. This query works on three datasets: NYTimes[4], LinkedMDB[5] and DBPedia[6]. The SPARQL endpoints of triple patterns for each dataset is specified using the SERVICE keyword.

The tuples of *S1* sub-query can be sent to both (*S1*, *S2*) and (*S1*, *S3*) joins, so that if the router sends them to (*S1*, *S3*) first, less intermediate results will be produced. On the other hand, if the source that sends the *S3* tuples has a latency, it will be better to change the plan. We generated a random value of latency for the *S3* source, with normal distribution up to a maximum latency value. We calculated the average percentage of tuples that were sent to (*S1*, *S3*) join for five times of running query. Then we increased the maximum latency of the *S1* source by 10ms, and followed up to a maximum latency of 150ms. The parameters of Algorithm 1 were fixed in all runs and set to *f*=0.1 and *d*=50ms.

---

[4] http://data.nytimes.com/
[5] http://linkedmdb.org/
[6] http://dbpedia.org/

```
SELECT * WHERE {
SERVICE <http://192.168.1.2:8890/sparql>{                                   S1
        ?dbfilm a dbpedia-owl:Film.
        ?dbfilm dbpedia-owl:budget ?budget .
        ?dbfilm dbpedia-owl:director ?director.}
SERVICE <http://192.168.1.3:8890/sparql>{                                   S2
        ?film a linkedMDB:film .
        ?film owl:sameAs ?dbfilm.
        ?film linkedMDB:genre <http://data.linkedmdb.org/resource/film_genre/4>.
        ?film linkedMDB:actor ?actor.}
SERVICE <http://192.168.1.4:8890/sparql>{                                   S3
        ?ydirector owl:sameAs ?director.
        ?ydirector nytimes:topicPage ?newspage.}}
```

**Figure 13:** The script for Query 1.

As can be seen in Figure 14, by increasing the latency of the *S3* source, the percentage of tuples that the router has sent them first to the (*S1*, *S3*) join is decreased. So we can conclude that the router has changed the plan so that it is not blocked and the *S1* tuples are first sent to the (*S1*, *S2*) join.

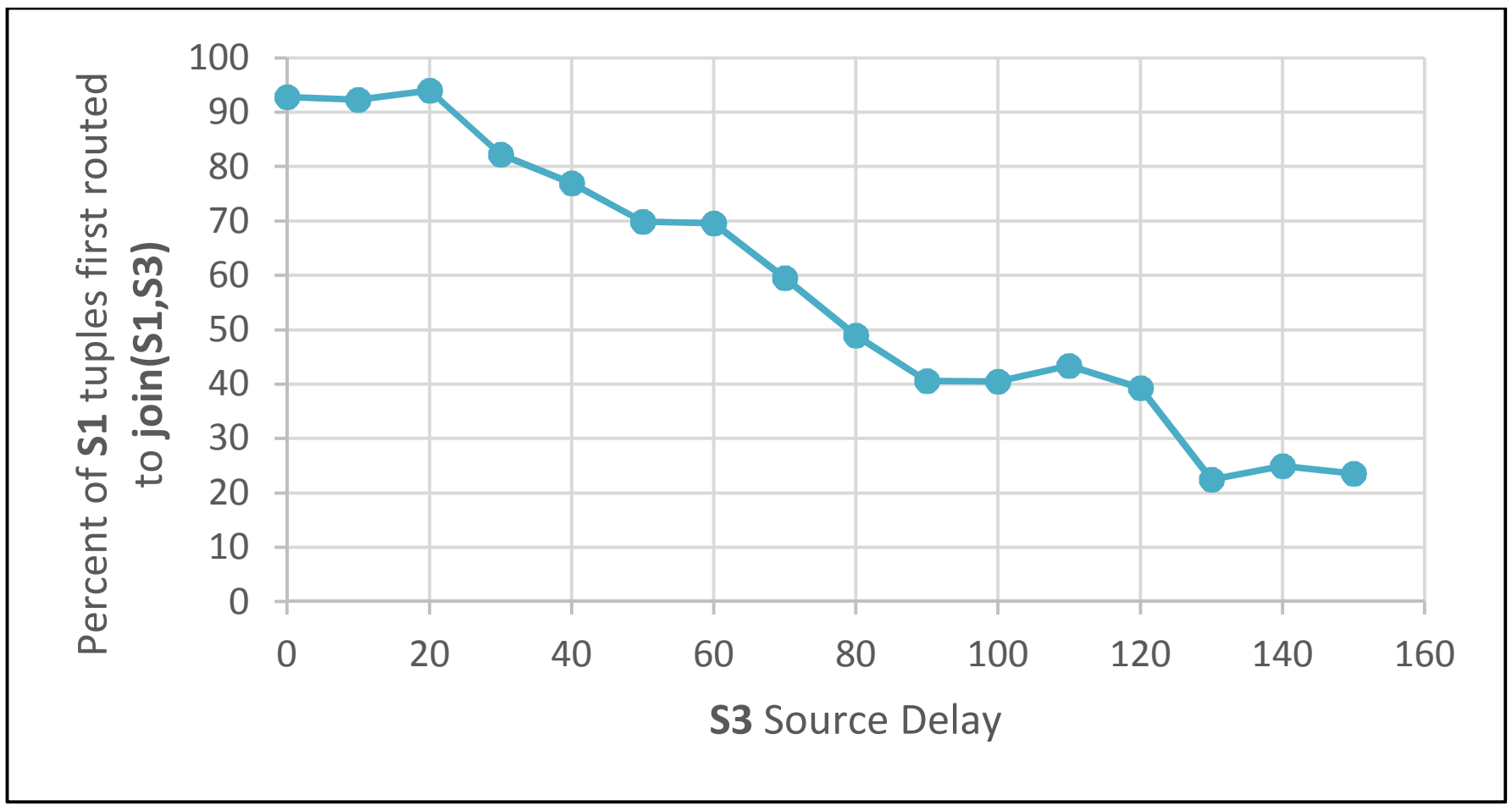

**Figure 14:** The results of evaluating the proposed routing strategy using the query in Figure 13

## *4.3 Comparing with Other Federation Engines*

In this subsection we compare our ADQUEX federation engine with FedX [10], SPLENDID [15], Jena ARQ[7] and ANAPSID [32] in terms of query execution times. For this comparison we used the CD and LS queries of the FedBench benchmark [37] plus six more queries that we wrote. We used these six queries because they have more joins that FedBench queries and their number of final results is also higher. For the details of these queries please refer to Appendix 1.

We converted the queries to SPARQL 1.1 and by specifying the sources using the SERVICE keyword. We ran these SPARQL 1.1 queries on our system as well as other systems that support SPARQL 1.1. For the systems that do not support the SERVICE keyword, we used SPARQL 1.0 queries. To correctly compare between our system which does not perform source selection, and systems that perform this step, we have deduced the source selection time and decomposition time from the total query execution time for these systems. The specification of the queries are summarized in Table 4.

[7] http://jena.apache.org/documentation/query/index.html

**Table 4:** The properties of queries used for evaluation. #P is the number of triple patterns, #J is the number of joins that the federation engine must perform if query correctly decomposed to sub-queries, and #R is the number of query results.

| Cross Domain (CD) | | | | Life Science (LS) | | | | Complex | | | |
|---|---|---|---|---|---|---|---|---|---|---|---|
| # | #P | #J | #R | # | #P | #J | #R | # | #P | #J | #R |
| **CD1** | 3 | - | 90 | **LS1** | 2 | - | 1159 | **C1** | 6 | 2 | 676 |
| **CD2** | 3 | 1 | 1 | **LS2** | 3 | - | 319 | **C2** | 8 | 2 | 2391 |
| **CD3** | 5 | 1 | 2 | **LS3** | 5 | 1 | 9054 | **C3** | 9 | 2 | 907 |
| **CD4** | 5 | 1 | 1 | **LS4** | 7 | 1 | 3 | **C4** | 8 | 2 | 358 |
| **CD5** | 4 | 1 | 2 | **LS5** | 6 | 2 | 393 | **C5** | 12 | 2 | 4749 |
| **CD6** | 4 | 1 | 11 | **LS6** | 5 | 1 | 28 | **C6** | 4 | 1 | 9126 |
| **CD7** | 4 | 1 | 1 | **LS7** | 5 | 1 | 1620 | | | | |

The execution times of the CD, LS and Complex groups of queries on different federation engines are shown in Figure 15, Figure 16 and Figure 17 respectively. The time-out value in this evaluation was set to five minutes and the value of "-1" in the tables is used to show that the query was timed out. As you can observe in Figure 15, our method had a higher execution time than other methods, for all of the queries of the CD group except for CD6 and CD7. For these two queries, the execution time was almost equal to that of FedX, and this time was better than other three systems. For queries of the LS group, as can be seen in Figure 16, although our method does not show the best execution time, but has performed well in most cases with acceptable execution times.

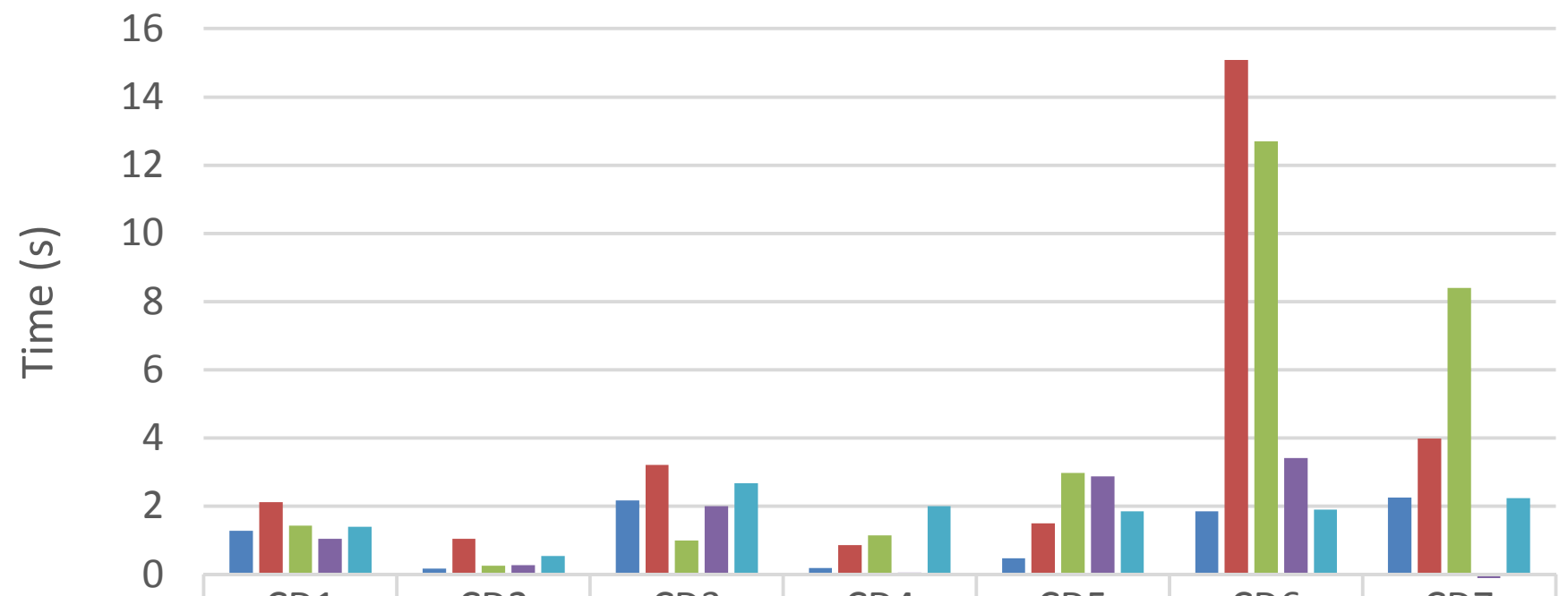

| | CD1 | CD2 | CD3 | CD4 | CD5 | CD6 | CD7 |
|---|---|---|---|---|---|---|---|
| FedX | 1.28 | 0.178 | 2.179 | 0.194 | 0.471 | 1.846 | 2.259 |
| ARQ | 2.127 | 1.049 | 3.208 | 0.86 | 1.502 | 15.079 | 3.98 |
| SPLENDID | 1.429 | 0.262 | 1.006 | 1.141 | 2.982 | 12.702 | 8.395 |
| ANAPSID | 1.043 | 0.271 | 1.998 | 0.064 | 2.87 | 3.413 | -1 |
| ADQUEX | 1.401 | 0.54 | 2.674 | 2.008 | 1.857 | 1.908 | 2.238 |

**Figure 15:** Comparison of the results of executing CD group of queries.

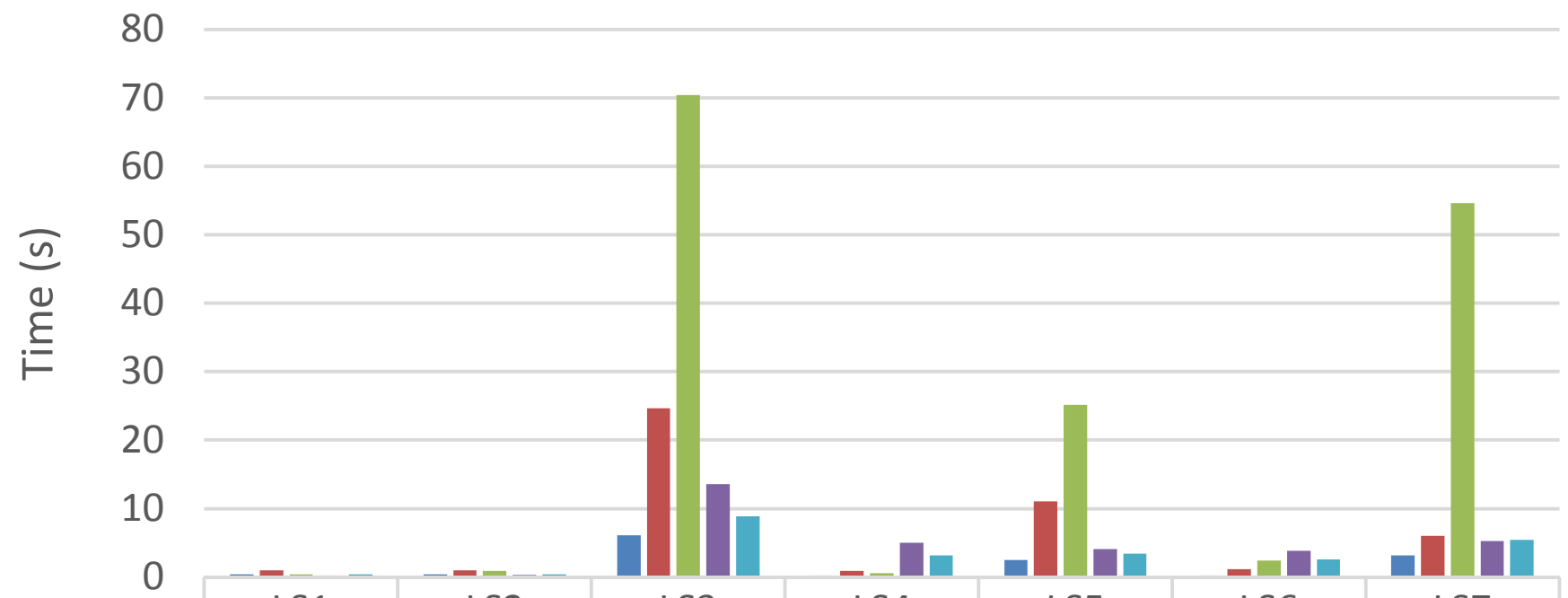

| | LS1 | LS2 | LS3 | LS4 | LS5 | LS6 | LS7 |
|---|---|---|---|---|---|---|---|
| FedX | 0.391 | 0.403 | 6.115 | 0.224 | 2.453 | 0.217 | 3.202 |
| ARQ | 1.007 | 0.942 | 24.652 | 0.928 | 11.018 | 1.131 | 6.028 |
| SPLENDID | 0.393 | 0.886 | 70.44 | 0.582 | 25.129 | 2.432 | 54.584 |
| ANAPSID | 0.204 | 0.265 | 13.6 | 4.984 | 4.074 | 3.824 | 5.237 |
| ADQUEX | 0.393 | 0.349 | 8.847 | 3.181 | 3.407 | 2.548 | 5.392 |

**Figure 16:** Comparison of the results of executing LS group of queries.

The queries in Complex group have a higher number of joins with more results. As can be seen in Figure 17, our method as well as ANAPSID performs much better than other methods for such type of queries. Moreover, our system has performed better than ANAPSID with a difference of multiple seconds.

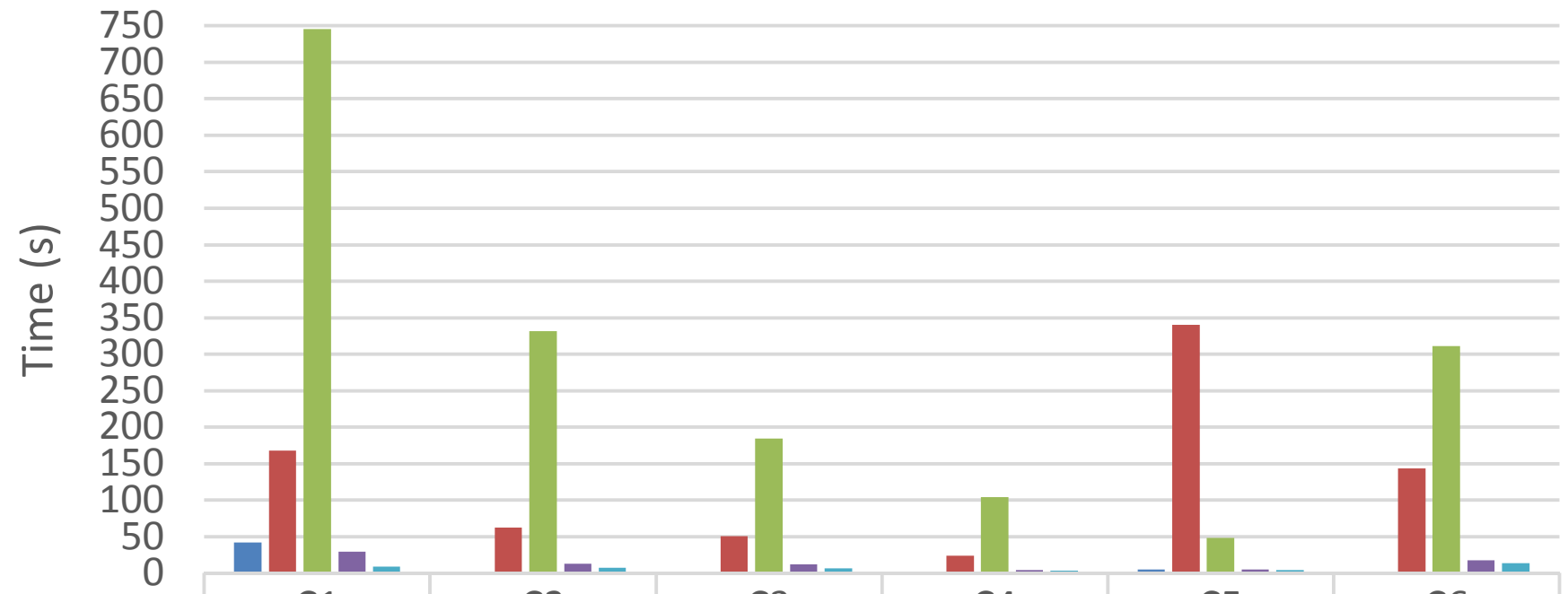

| | C1 | C2 | C3 | C4 | C5 | C6 |
|---|---|---|---|---|---|---|
| FedX | 42.246 | -1 | -1 | -1 | 5.346 | -1 |
| ARQ | 168.035 | 62.518 | 50.776 | 23.839 | 339.91 | 143.233 |
| SPLENDID | 745.49 | 331.428 | 184.782 | 104.112 | 48.236 | 311.328 |
| ANAPSID | 29.68 | 13.155 | 11.748 | 4.484 | 5.062 | 17.792 |
| ADQUEX | 9.044 | 7.135 | 6.845 | 3.428 | 4.014 | 13.372 |

**Figure 17:** Comparison of the results of executing Complex group of queries.

## *4.4 Interpretation of the results*

By observing both the response time and the specification of the queries, we can conclude that our ADQUEX method performs better than existing methods in situations that one of the following conditions hold:

1) The number of joins in a query is more than one

2) If a query has, for example, two sub-queries then the number of results of one sub-query is not very low while the other's is very high.

For most of the queries in CD group, none of the above conditions hold. As you can see in Table 4, all of the queries in this group have a join on the federation engine side. All queries in this group have two sub-queries and in most cases a sub-query has very few results while another sub-query has a relatively high number of results. In such situations, using the bind join operator is useful for join, since it can reduce the network cost. On the contrary, we have used SHJ, hence the difference in execution time.

For the queries of the LS group both of the above conditions hold, and as you can see that ADQUEX has performed well for this group. The queries of the Complex group have more joins and produce more results than the queries of the CD and LS group, and they have no queries with extremely low number of results. Therefore, using the bind join for such queries will increase the response time of the SPARQL endpoints which in turn leads to increased query execution time. In such situations, using SHJ will perform better for join operations, as applied in our method.

# 5 Conclusion

In this paper, we offered a novel solution for processing the federated SPARQL queries in an adaptive manner. The proposed ADQUEX system is able to run federated queries that are defined by standard SPARQL 1.1 language, on-demand, over a federation of SPARQL endpoints without any need for preprocessing to gather statistics. The proposed query executor uses a router at its core, and the router exploits a routing strategy to find the best path for sending the tuples. We presented a new routing strategy and evaluated it in different scenarios. The results show that the proposed routing strategy can find the best plan for query execution at runtime. The proposed routing strategy has a high stability as it can find the best plan in most cases.

In our evaluations, the proposed system was run with complex queries and real-world data sources in a real network environment. It was compared with FedX, SPLENDID and Jena ARQ query engines in terms of query execution time. The results show that the execution time of our system is very close to the aforementioned engines based on FedBench benchmark queries and performs faster than other engines for some queries. The performance of the proposed system was much higher than other engines for larger queries with more data, showing a great performance improvement. We can conclude that our method performs faster for large queries, but may be slower for small queries with a few number of results because of not selecting the proper join operator in smaller cases.

We plan to extend our method in future so that in addition to the optimized order of operators, the physical operators being also selected adaptively according to the properties of the query and runtime environment conditions. This will improve the execution time in almost every runtime situation.